\documentclass[12pt]{JHEP3}
\usepackage{amsmath,epsf,amssymb,latexsym,cite,graphics,bbm}

\DeclareFontFamily{U}{rsf}{}
\DeclareFontShape{U}{rsf}{m}{n}{
  <5> <6> rsfs5 <7> <8> <9> rsfs7 <10-> rsfs10}{}
\DeclareMathAlphabet\Scr{U}{rsf}{m}{n}

\def\O{\Scr{O}}

\def\Z{\Scr{H}}

\def\C{{\mathbb C}}

\def\P{{\mathbb P}}
\def\R{{\mathbb R}}
\def\Z{{\mathbb Z}}

\def\End{\operatorname{End}}

\def\Res{\operatorname{Res}}

\def\GL{\operatorname{GL}}

\def\GU{\operatorname{U{}}}

\def\Hess{\operatorname{Hess}}

\def\diag{\operatorname{diag}}

\def\p{\partial}

\def\cD{{\Scr D}}

\def\cL{{\Scr L}}

\def\ff#1#2{{\textstyle\frac{#1}{#2}}}

\def\ep{\epsilon}

\def\thetab{\overline{\theta}}

\def\la{\langle}
\def\ra{\rangle}
\def\lad{\langle\!\langle}
\def\rad{\rangle\!\rangle}

\def\Upsilonb{\overline{\Upsilon}}

\def\Jt{{\widetilde{J}}}

\def\Wt{{\widetilde{W}}}

\def\Sigmab{\overline{\Sigma}}

\def\Phib{\overline{\Phi}}

\def\cDb{\overline{\cD}}
\def\Gammab{\overline{\Gamma}}
\def\Upsilonb{\overline{\Upsilon}}

\title{Half-Twisted Correlators from the Coulomb Branch}
\author {Jock McOrist and Ilarion V. Melnikov \\
\normalsize Enrico Fermi Institute \\
\normalsize University of Chicago \\
\normalsize Chicago, IL 60637, USA\\
Email:  \email{jmcorist@uchicago.edu,lmel@theory.uchicago.edu}
}
\abstract{We compute correlators of chiral operators in half-twisted $(0,2)$ supersymmetric gauged linear sigma models.
Our results give simple algebraic formulas for a $(0,2)$ generalization of genus zero Gromov-Witten invariants
of compact toric varieties.  We derive compact expressions for $(0,2)$ deformed quantum cohomology relations and
apply our general method to several examples.}

\preprint{EFI-07-40}

\begin{document}

%%%%%%%%%%%%%%%%%%%%%%%%%%%%%%%%%%%%%%%%%%%%%%%%%%%%%%%%%

\section{Introduction}

It is well known that the $A$-topological twist of a $d=2$, $N=2$ SUSY theory
provides a quick route towards the
computation of correlators of chiral observables.  These correlators are physically interesting because while they probe the IR
structure of the theory, the topological twist renders them computable in terms of UV variables.  When specialized to $N=2$
sigma models with target-space $M$, these topological correlators compute genus zero Gromov-Witten invariants of $M$.

The models of most interest to physicists and mathematicians alike are those where $M$ is a Calabi-Yau three-fold.  These
theories flow to a non-trivial superconformal field theory which may be used for string compactification, and the
corresponding manifolds possess a rich enumerative structure.  It is often the case that computations in these
conformal theories may be related to computations in related gapped models.  For example, correlators in the quintic
three-fold in $\P^4$ may be determined in terms of correlators of the $\P^4$ sigma model.  A physical realization of
this relation was given in~\cite{Morrison:1994fr} by using  the gauged linear sigma model (GLSM) introduced
in~\cite{Witten:1993yc}.

The $A$-twisted GLSM is not only useful in relating the conformal and gapped theories, but it also provides a simpler way
to solve the gapped models.  For example, even in the ``simple'' case of $\P^4$ the computation in the non-linear sigma model
description requires a careful compactification of the moduli space of world-sheet instantons.  In the linear model, the
corresponding computations involve sums of gauge instantons with compact and toric moduli spaces.
Thus, the instanton sums, and hence correlators, are computable by combinatoric techniques.

In the case of compact and toric GLSMs, there is an additional simplification:  one may reduce the instanton sums to simpler
computations in Landau-Ginzburg theories.  This may be done by carrying out an abelian duality as in \cite{Hori:2000kt}, or
by working on the Coulomb branch of the GLSM \cite{Melnikov:2006kb}.  The former requires a careful analysis of the map to
dual variables and the associated Jacobian factors in the path integral measure, while the latter is stated in terms of the original
fields and parameters of the GLSM.

Much of the structure just described does not require $(2,2)$ supersymmetry.  In particular,
in a large class of models $(0,2)$ preserving deformations away from the $(2,2)$ locus are unobstructed~\cite{Distler:1986wm, Silverstein:1995re,Basu:2003bq,Beasley:2003fx}.  While these deformations deform the $(2,2)$ chiral ring structure,
they do not destroy it~\cite{Adams:2003zy, Sharpe:2005fd, Adams:2005tc}.  Once
the theory is deformed from the $(2,2)$ locus, it no longer admits a topological twist.  However, under favorable circumstances,
it is still possible to perform a half-twist~\cite{Witten:1991zz}.  Although no longer topological, the half-twisted theory is still sufficiently
simple to render correlators of local chiral operators readily computable~\cite{Melnikov:2007xi}.

The GLSM continues to be of great service even off the $(2,2)$ locus.  First, as already observed in~\cite{Witten:1993yc}, the linear
model provides a simple presentation of the $(0,2)$ deformations.  On the $(2,2)$ locus the left-moving fermions couple to the
tangent bundle of the variety, and the GLSM Lagrangian neatly separates the deformations of the tangent bundle of the ambient toric
variety from deformations associated to the choice of hypersurface/complete intersection.  Second, by working with the GLSM a $(0,2)$
generalization of the abelian duality of Hori and Vafa was derived in~\cite{Adams:2003zy}.  Finally, the structure of the gauge-instanton
moduli spaces is still simple enough that direct computations of correlators are possible~\cite{Katz:2004nn,Guffin:2007mp}.

In this paper we will compute genus zero correlators of chiral observables in $(0,2)$-deformed compact and toric GLSMs by working on the
Coulomb branch of the theory.  Our method yields simple algebraic expressions for the amplitudes and leads to formulas for
the deformed quantum cohomology rings.  The derivation is a straight-forward generalization of $(2,2)$ Coulomb branch techniques
and uses some recent results on correlators in half-twisted $(0,2)$ Landau-Ginzburg theories.

The linear model computations are sure to play an important role in physics and mathematics.  We expect the explicit form of these
correlators to be useful in generalizing mirror symmetry, special geometry, and  Gromov-Witten invariants. It is likely that these
amplitudes will allow us to compute Yukawa couplings in a large class of $(0,2)$ heterotic compactifications.

The rest of the paper is organized as follows.  We describe the $(0,2)$-deformed compact and toric GLSMs in section~\ref{s:glsmagain}.
In section~\ref{s:potential}, we compute the effective potential describing the Coulomb vacua, and we use it to arrive at a general
formula for half-twisted correlators in section~\ref{s:correlators}.  We present two massive examples in section~\ref{s:gexamples}
and a compact conformal example in section~\ref{s:hyperexample}.  We conclude in section~\ref{s:conclusions}.  The appendix contains an
example of a simple Maple code to compute correlators in the theory studied in~\cite{Adams:2003zy,Katz:2004nn,Guffin:2007mp}.

%%%%%%%%%%%%%%%%%%%%%%%%%%%%%%%%%%%%%%%
\section{A Brief Review of $(0,2)$ Linear Models} \label{s:glsmagain}
In this section we review the Lagrangian of $(0,2)$ deformations of a $(2,2)$ linear model in standard $(0,2)$ superspace notation~\cite{Witten:1993yc}.
As this is well known material, we will not present the component expansions of the superfields.
\subsection{The $(2,2)$ Theory}
We will denote the $(0,2)$ superspace derivatives by
$\cD_+, \cDb_+$.  The $(2,2)$ linear model is an abelian gauge theory with matter multiplets $(\Phi^i, \Gamma^i)$, $i=1,\ldots,n$
coupled to vector multiplets $V_{a,\pm}$ with integral charges $Q^a_i$, $a=1,\ldots,n-d$.  In addition, the theory contains $n-d$
neutral multiplets $\Sigma_a$. The $\Phi^i$ and $\Sigma_a$ are chiral
bosonic multiplets satisfying $\cDb_+ \Phi^i = 0$, while the $\Gamma^i$ are
fermionic multiplets with $\cDb_+ \Gamma^i = \sqrt{2}E^i(\Phi,\Sigma)$.   The gauge field-strengths corresponding to the $V_{a,\pm}$ live in chiral
fermionic multiplets $\Upsilon_a$.  With this field content, the Lagrangian takes the form
$S = S_{\text{kin}}+ S_{\text{F-I}}+S_J$, with
\begin{eqnarray}
\label{eq:GLSMaction}
S_{\text{kin}} & = & \int d^2 y d^2\theta \left\{ - \ff{1}{8 e_0^2} \Upsilonb_a \Upsilon_a -\ff{i}{2e_0^2} \Sigmab_a \p_-\Sigma_a
- \ff{i}{2} \Phib^i (\p_-+iQ^a_i V_{a,-}) \Phi^i - \ff{1}{2} \Gammab^i \Gamma^i\right\}, \nonumber\\
S_{\text{F-I}} & = & \ff{1}{8\pi i} \int d^2y d\theta^+  \Upsilon_a \log(q_a) |_{\thetab^+ = 0} + \text{h.c.}, \nonumber\\
S_J & = &  \int d^2 y d\theta^+ \Gamma^i J_{i} (\Phi)|_{\thetab^+ =0} + \text{h.c.}.
\end{eqnarray}
The $q_a = e^{-2\pi r_a + i \theta_a}$ parametrize the Fayet-Iliopulos terms (the $r^a$) and the $\theta$-angles of the gauge theory,
while the $J_i(\Phi^i)$ are polynomials with charges $-Q^a_i$.  The action is $(2,2)$ supersymmetric when $J_i = \p W/\p\Phi^i$ for
some gauge-invariant superpotential $W$, and
\begin{equation}
E^i = i \sqrt{2} \Sigma_a Q^a_i \Phi^i.
\end{equation}
More generally, the theory has $(0,2)$ supersymmetry as long as $\sum_i E^i J_i = 0$.
In what follows, we will mostly consider theories with $J_i = 0$.
We will refer to such GLSMs as {\em toric}, because for generic values of the $r^a$ the classical bosonic moduli space is a toric
variety.  We may always choose a basis for the gauge charges so that when $r^a \gg 0$  for all $a$ the classical moduli space is a smooth toric
variety $X$ of dimension $d$, and at low energies the GLSM is well-described by a non-linear sigma model with target-space $X$.  When $X$ is
compact, we will say the corresponding GLSM is compact.\footnote{The reader will find a more precise discussion of this terminology
in~\cite{Melnikov:2006kb}.}

\subsection{$(0,2)$ Deformations}
We now wish to consider $(0,2)$ deformations of compact, toric $(2,2)$ GLSMs.  These deformations are obtained by taking more
general forms of the $E^i(\Sigma,\Phi^i)$.  In this paper we will consider $E^i$ that remain linear in the $\Sigma_a$ and $\Phi^i$:
\begin{equation}
E^i(\Sigma,\Phi) = i \sqrt{2} \Sigma_a \left(A^a\right)^{i}_{j} \Phi^j,
\end{equation}
where $A^{ai}_{j}$ is an array of $n^2(n-d)$ complex parameters.\footnote{There are good reasons for restricting to this form of the $E^i$.
Terms of higher order in the $\Sigma_a$ will typically lead to additional vacua in the geometric phase, while terms of higher order
in the $\Phi^i$ will not affect correlators as long as large generic $\Sigma$ VeVs give masses to all the matter multiplets.}

The $A^{ai}_j$ are constrained by gauge invariance.  Since the $\Gamma^i$ and $\Phi^i$ have identical gauge charges, the
$A$ may only mix fields that have identical gauge charges for all gauge groups.  We will keep track of this by partitioning
the $(\Phi^i,\Gamma^i)$ into sets with identical charges:
\begin{equation}
\{\Phi^i,~~i=1,\ldots n\} \to \cup_\alpha \{ \Phi_{(\alpha)}^{I_\alpha},~~ I_\alpha=1,\ldots, n_\alpha\},
\end{equation}
with $\sum_\alpha n_\alpha = n$, and $Q^a_{I_\alpha} = Q^a_{J_\alpha}=Q^a_{(\alpha)}$ for all $a,\alpha$ and $I_\alpha,J_\alpha$.
We then have
\begin{equation}
E_{(\alpha)}^{I_\alpha} = i\sqrt{2}~ \sum_{a=1}^{n-d} \Sigma_a \left[A_{(\alpha)}^a\right]^{I_\alpha}_{J_\alpha} \Phi_{(\alpha)}^{J_\alpha}.
\end{equation}
Not all parameters in the $A_{(\alpha)}$ correspond to deformations of the theory.  As we will see below, a number of these
may be absorbed into field re-definitions.  In what follows, we will suppress the $I_\alpha, J_\alpha$ indices whenever it is
unlikely to cause confusion, and we will find it useful to work with the $n_\alpha\times n_\alpha$ matrices
\begin{equation}
\label{eq:M}
M_{(\alpha)}(\Sigma)  =  \sum_{a=1}^{n-d} \Sigma_a A_{(\alpha)}^{a},
\end{equation}
as well as vectors
\begin{equation}
\Phi_{(\alpha)} = ~^{~~t}(\Phi_{(\alpha)}^{1}, \ldots, \Phi_{(\alpha)}^{n_\alpha}).
\end{equation}
The bosonic potential that follows from the action takes the form
\begin{equation}
\label{eq:Uclass}
U = 2\sum_{\alpha} \phi_{(\alpha)}^\dag M_{(\alpha)}^\dag M_{(\alpha)} \phi_{(\alpha)}
      + \ff{e_0^2}{2} \sum_{a=1}^{n-d} \left(\sum_\alpha Q^a_{(\alpha)} \phi^\dag_{(\alpha)} \phi_{(\alpha)} - r^a\right)^2.
\end{equation}

%%%%%%%%%%%%%%%%%%%%%%%%%%%%%%%%%%%%%%%%%%%%%%%%
\section{The Effective Potential on the Coulomb Branch} \label{s:potential}
Consider the classical parameter space of a compact, toric GLSM described above.  Ignoring the $\theta$-angles, this is just
the space $\R^{n-d}$ corresponding to the $n-d$ Fayet-Iliopulos terms $r^a$.   Let $K_c \subset \R^{n-d}$  be the cone generated
by the $n$ vectors $Q_i \in \R^{n-d}$:
\begin{equation}
K_c = \{r^a = \sum_i Q^a_i \xi^i ~~| ~~\xi \in \R_{\ge 0}^{n} \}.
\end{equation}
When the $r^a \in K_c$, and the parameters in the matrices $M_{(\alpha)}$ are generic, the $\sigma_a$ fields are massive, and the
classical moduli space of the GLSM is a toric variety of dimension $d$.  In general, $K_c$ consists of a number of subcones corresponding
to various geometric ``phases'' of the GLSM.  When $r^a \not\in K_c$, and the $M_{(\alpha)}$
are generic, there are no classical supersymmetric vacua.  Nevertheless, supersymmetry is unbroken for parameters outside of $K_c$.
\footnote{For example, considerations of topological invariants such as the Witten index, suggest that the theory should possess SUSY
vacua for all $r^a$.}   In
the case of $(2,2)$ models~\cite{Witten:1993yc,Morrison:1994fr}, the supersymmetric ground states in this region of parameter space
are  massive Coulomb vacua. These are charcterized by large $\sigma_a$ VeVs, which give large masses to the matter multiplets
$(\Phi^i,\Gamma^i)$.  The dynamics of the $(\Sigma,\Upsilon)$ fields are governed by an effective twisted superpotential $\Wt(\Sigma)$,
which in $(0,2)$ language takes the form
\begin{equation}
\cL_{\text{eff}} =  \int d\theta^+ \left. \Upsilon_a \frac{\p \Wt}{\p\Sigma_a}\right|_{\thetab^+=0} + \text{h.c.} ~.
\end{equation}
For generic values of the parameters, this interaction gives masses to all $(\Sigma_a,\Upsilon_a)$ multiplets.
This potential is one-loop exact, as may be seen by 't Hooft anomaly matching, and it is self-consistent\footnote{In other words, it predicts large
$\sigma_a$ VeVs and thus large masses for the $(\Phi^i,\Gamma^i)$ multiplets.} when the $r^a$ are deep in the ``non-geometric'' phase.

This result is easily generalized off the $(2,2)$ locus.  Provided that the $M_{(\alpha)}$ are chosen so that non-zero
$\sigma_a$ VeVs give masses to all the matter fields (this will be true for small deformations off the $(2,2)$ locus),  the one-loop shift in the $D$-term tadpole is given by
\begin{equation}
\delta \la -\ff{1}{e_0^2} D_a \ra = \sum_{\alpha} Q_a^{(\alpha)} \sum_{I_{\alpha}=1}^{n_\alpha}
\int \frac{d^2 k}{(2\pi)^2}\left\{ \frac{1}{k^2+ 2 m_{(\alpha) I_\alpha}^2} - \frac{1}{k^2+2\mu^2}\right\}.
\end{equation}
Here the $m_{(\alpha) I_\alpha}^2$ are the positive eigenvalues of the mass matrix $M_{(\alpha)}^\dag M_{(\alpha)}$, and
$\mu$ is a subtraction point whose choice may be absorbed into a renormalization of the Fayet-Iliopulos parameters $r^a$.

Carrying out the integral, we find a shift that may be interpreted as a $(0,2)$ potential
\begin{equation}
\cL_{\text{eff}} =   \int d \theta^+ \sum_{a=1}^{n-d}\Upsilon_a \Jt_a(\Sigma)|_{\thetab^+=0} + \text{h.c.},
\end{equation}
with
\begin{equation}
\label{eq:J}
\Jt_a  = -\frac{1}{8\pi i}  \log \left[
\prod_\alpha \left(\frac{\det M_{(\alpha)}}{\mu^{n_\alpha}}\right)^{Q_{(\alpha)}^a} / q_a(\mu)\right].
\end{equation}
Just as on the $(2,2)$ locus, this potential is 1-loop exact. The massive Coulomb vacua are common zeroes of $\Jt_a(\sigma) = 0$,
i.e. the $\sigma_a$ satisfying
\begin{equation}
\label{eq:Jvacua}
\prod_{\alpha} \left(   \frac{\det M_{(\alpha)}(\sigma)}{\mu^{n_\alpha}} \right)^{Q^a_{(\alpha)}} = q_a(\mu).
\end{equation}
It is easy to see that on the $(2,2)$ locus the $\Jt_a$ derived above follow from the effective twisted superpotential
of~\cite{Witten:1993yc,Morrison:1994fr}.

\section{Correlators in the Half-Twisted Model}\label{s:correlators}
A toric GLSM on the $(2,2)$ locus possesses two classical $\GU(1)$ symmetries---the left- and right-moving R-symmetries,
which act with charges
\begin{equation}
\begin{array}{|c|c|c|c|c|c|}\hline
\ast 		& \theta^+ 	& \Phi^i & \Gamma^i 	& \Sigma_a 	& \Upsilon_a \\ \hline
\GU(1)_R  &    1         	&    0     &     0               	&   1       		& 1 		 \\ \hline
\GU(1)_L 	&     0		&    0     &      -1             	&   -1      		& 0 		 \\ \hline
\end{array}
\end{equation}
The vectorial combination of the corresponding currents, $J_V = J_R+J_L$ is non-anomalous and may be used to
twist the theory~\cite{Witten:1991zz,Witten:1993yc}.  This is the standard $A$-twist of the linear model---a
topological field theory.  This theory is endowed with a nilpotent fermionic symmetry generated by  BRST-like operator
$Q_A$.  The (local) observables of the theory are local, gauge-invariant operators in the $Q_A$ cohomology.  In
the linear model these are given by the $\sigma_a(x)$.  The $A$-model correlators are just the genus zero
 amplitudes $\la \sigma_{a_1}(x_1) \cdots \sigma_{a_k}(x_k) \ra$.

The conservation of the current $J_V$ is preserved by $(0,2)$ deformations of a toric GLSM, and hence it may
still be used to twist the theory\cite{Witten:1991zz}.  The resulting half-twisted theory, while no longer topological, is
also endowed with a BRST-like operator, $Q_T$.  Unlike the cohomology of $Q_A$, the cohomology of $Q_T$ is
in general infinite-dimensional.  Nevertheless, it has a meaningful truncation to a finite-dimensional ``zero-energy''
sub-space, which on the $(2,2)$ locus matches the cohomology of $Q_A$\cite{Adams:2005tc}.  Thus, even in
the half-twisted model it is interesting to compute the correlators of the $\sigma_a(x)$.

In the remainder of this section we will argue that in a compact toric half-twisted linear model these amplitudes are
given by
\begin{equation}
\label{eq:famous_result}
\la\sigma_{a_1}(x_1) \cdots \sigma_{a_k}(x_k) \ra = \sum_{\sigma|\Jt(\sigma)=0}
\sigma_{a_1} \cdots \sigma_{a_k} \left[ \det_{a,b} (\Jt_{a,b}) \prod_\alpha \det M_{(\alpha)} \right]^{-1},
\end{equation}
with $M_{(\alpha)}$ defined in eqn.~(\ref{eq:M}), and the $\Jt_a$ given in eqn.~(\ref{eq:J}).
The result follows from a combination of  observations on correlators in massive half-twisted Landau-Ginzburg
theories~\cite{Melnikov:2007xi} and $A$-model computations on the Coulomb branch on the $(2,2)$
locus~\cite{Melnikov:2006kb}.

The starting point for the argument is the observation that a constant rescaling of the world-sheet metric is a $Q_T$-exact
deformation of the action. Since $Q_T$-exact operators decouple from $Q_T$-closed operators, the amplitudes are
independent of such a rescaling. In the limit of a large world-sheet,  it is clear that the correlators in these massive
theories are independent of the positions $x_k$ and may be computed exactly by a semi-classical expansion.

In addition, just as on the $(2,2)$ locus, we expect\footnote{Our findings will confirm this expectation.} the correlators to
be meromorphic functions of the linear model parameters, so that the  result of a semi-classical expansion in any phase of
the linear model will be easy to continue is to any other phase.  While the answers obtained in the various phases are
simply related, the degree of computational complexity changes significantly depending on which phase is used.

\subsection{Computations in a Geometric Phase}
In the geometric phases, i.e. when the $r^a$ are chosen to lie in $K_c$, the relevant semi-classical expansion is in terms of the
gauge instantons of the linear model.  On general grounds, we expect the correlators to be of the form
\begin{equation}
\la \sigma_{a_1}(x_1) \cdots \sigma_{a_k}(x_k) \ra = \sum_{N} c_N(A) q^N,
\end{equation}
where $N$ is a multi-index $N_1,\ldots,N_{n-d}$ labelling elements of $H_2(M,\Z)$, and the $c_N(A)$ are coefficients that
depend on the $(0,2)$ deformation parameters.  This may be thought of as a $(0,2)$ generalization of Gromov-Witten theory \cite{Katz:2004nn}.
These sums have been explored in great detail on the $(2,2)$ locus~\cite{Cox:2000vi,Hori:2003ic}.  More recently,
the $c_N(A)$  were computed  in a $(0,2)$ deformed linear model for $\P^1\times \P^1$ in \cite{Katz:2004nn,Guffin:2007mp}.
The computation off the $(2,2)$ locus is much more involved, but despite the complexity that arises in the intermediate steps, the final
results are elegant and compact expressions for the correlators.

\subsection{Correlators in the Non-Geometric Phase}
In $(2,2)$ models the computations in the non-geometric phase, i.e. when the $r^a \not \in K_c$, are considerably simpler than the
geometric phase instanton sums.  It was shown in~\cite{Melnikov:2006kb} that the correlators take the form
\begin{equation}
\la \sigma_{a_1}(x_1) \cdots \sigma_{a_k}(x_k) \ra = \sum_{\sigma|d\Wt(\sigma)=0} \sigma_{a_1} \cdots \sigma_{a_k}
\left[ \det \Hess \Wt(\sigma) \prod_i (Q^b_i \sigma_{b} ) \right]^{-1},
\end{equation}
where $\Wt$ is the one-loop twisted effective superpotential.  There is a simple way to understand this formula.
In the non-geometric phase the semi-classical field configurations are given by $\phi^i=0$ and $\sigma_a$ fixed to the critical
points of $\Wt(\sigma)$.  Expanding the action in fluctuations about one of the critical points, we find that the integration over
the $\Sigma_a,\Upsilon_a$ fields leads to the usual Landau-Ginzburg contribution of $\det\Hess\Wt^{-1}$\cite{Vafa:1990mu}.
The integration over the zero modes of the $(\Phi^i, \Gamma^i)$ multiplets produces an additional contribution of
$\prod_i (Q^b_i \sigma_b)^{-1}$.

In contrast to the geometric phase analysis, the computation in the non-geometric phase is not much more involved off the $(2,2)$
locus.  The semi-classical field configurations are given by $\phi^i =0$ and $\sigma_a$ fixed to the common zeroes of the $\Jt_a$
of eqn.~(\ref{eq:J}).  The expansion in fluctuations about these configurations is easily carried out. The fluctuations of the
$(\Upsilon_a,\Sigma_a)$ multiplets lead to the $\det \Jt^{-1}$ in the measure, while the zero modes of the $(\Phi^i,\Gamma^i)$ multiplets
lead to the additional factor of $\prod_\alpha \det M_{(\alpha)}^{-1}$.  The former contribution is familiar from $(0,2)$ Landau-Ginzburg theories
analyzed in~\cite{Melnikov:2007xi}. The latter arises from the $\sigma$-dependent mass term for the $\phi_{(\alpha)}$ in eqn.~(\ref{eq:Uclass})
and its supersymmetric completion.  Combining these contributions and summing over the common zeroes of the $\Jt_a$ leads to the
expression advertised in eqn.~(\ref{eq:famous_result}).

\section{Remarks on the Correlators}
We will now make several general remarks on the half-twisted correlators.  First, let us dispose of the mass scale $\mu$.  In the
untwisted theory, this RG scale describes the running of the parameters $q_a(\mu)$.  In the half-twisted theory there is no
longer any meaningful running, and as we restrict computations to zero energy correlators, $\mu$ is just a length-scale.  In what
follows, we will work in units of $\mu$.  Second, we will not attach a particular meaning to an over-all normalization constant of
the correlators, so will not keep track of constant factors like the $(8\pi i)^{-1}$ in the $\Jt_a$.

\subsection{$(0,2)$ Deformations of Quantum Cohomology}
It is easy to see that the correlators satisfy the relations
\begin{equation}
\label{eq:qcoho}
\la \O \prod_{\alpha| Q^a_{(\alpha)} > 0} \det M_{(\alpha)}^{Q^a_{(\alpha)}} \ra = q_a
\la \O \prod_{\alpha | Q^{a}_{(\alpha)} < 0} \det M_{(\alpha)}^{-Q^a_{(\alpha)}} \ra,~~~\text{for all}~~ \O~~\text{and}~~a.
\end{equation}
These relations are the $(0,2)$ deformed version of the usual quantum cohomology of the linear model.  They were discussed in
specific examples in \cite{Adams:2003zy,Katz:2004nn}.

These are powerful constraints on the correlators, which often determine most of the correlators in terms of a small finite
subset of amplitudes.  When interpreted in terms of computations in a smooth geometric phase these relations are a
quantum deformation of the usual cohomology ring of the toric variety.  We may always choose a basis of charges such
that the smooth geometric phase corresponds to sending all the $q_a \to 0$.  We see that in that limit the relations reduce
to a $(0,2)$-deformation of the Stanley-Reisner relations.

\subsection{The $M_{(\alpha)}$ and Bundle Deformations}
Another simple consequence of eqn.~(\ref{eq:famous_result}) is that our results are invariant under the transformations
\begin{equation}
M_{(\alpha)} \to U^{-1}_{(\alpha)} M_{(\alpha)} U_{(\alpha)},~~~U\in \GL(n_\alpha,\C).
\end{equation}
This suggests that not all parameters in the $M_{(\alpha)}$ correspond to genuine deformations of the theory.  This
is not a surprise: by working in a geometric phase, it is easy to see that the $E^i$
overparameterize the deformations of the tangent bundle of the toric variety~\cite{Guffin:2007mp}.

This over-parametrization is easily quantified in the linear model.  Recall that the $E^i$ enter the theory via the relation
\begin{equation}
\cDb_+ \Gamma^i = E^i,
\end{equation}
which we may equivalently write as
\begin{equation}
\cDb_+ \Gamma_{(\alpha)} = M_{(\alpha)} \Phi_{(\alpha)}.
\end{equation}
Thus, a similarity transformation $M_{(\alpha)} \to U^{-1}_{(\alpha)} M_{(\alpha)} U_{(\alpha)}$ may be absorbed into a field
re-definition
\begin{equation}
\Phi_{(\alpha)} \to U_{(\alpha)} \Phi_{(\alpha)},~~~\Gamma_{(\alpha)} \to U_{(\alpha)} \Gamma_{(\alpha)}.
\end{equation}
This allows us to eliminate $\sum_\alpha (n_\alpha^2-1)$ parameters from the $E^i$.  In addition, a change of
basis on the $\Sigma_a$ may be used to eliminate $(n-d)^2$ degrees of freedom in the $E^i$.  Thus, we find
that there should be
\begin{equation}
N_E = (n-d-1) \sum_\alpha n_\alpha^2 +\sum_\alpha 1 - (n-d)^2
\end{equation}
parameters that cannot be absorbed into field re-definitions.
When the linear model is in the smooth geometric phase, we expect these to correspond to deformations of the
tangent bundle.

%%%%%%%%%%%%%%%%%%%%%%%%%%%%%%%%%%%%%%%
\section{Compact, Toric Examples} \label{s:gexamples}

\subsection{$(0,2)$ deformations of $\P^1\times \P^1$}
The linear model for this theory has $n=4$, $n-d = 2$, with charges
\begin{equation}
Q = \left(
\begin{array}{ccccc}
  			     	1    &   1       & 0        &    0  \\
				0    &   0       & 1        &    1
\end{array}
\right).
\end{equation}

The $(2,2)$ locus is a GLSM for target-space $X = \P^1\times\P^1$.
The $(0,2)$ deformations are described by the matrices $M_{(1)}$ and $M_{(2)}$,
mixing the $\{\phi^1,\phi^2\}$, and $\{\phi^3,\phi^4\}$, respectively.  Taking the
redundancies described above into account, we expect $N_E = 6$ deformations.

This simplest example of a compact toric GLSM with $(0,2)$ deformations was considered
in~\cite{Adams:2003zy} and later studied in~\cite{Katz:2004nn}.  Recently, Guffin and Katz
computed the two-point and four-point functions in this theory, taking into account all the bundle
deformations~\cite{Guffin:2007mp}.  A computation of $H^{1}(X,\End{TX})$ shows that there
are six $(0,2)$ deformations of this theory, in agreement with the count above.  We will
parametrize the $(0,2)$ deformations in the same fashion as in~\cite{Guffin:2007mp}.  Introducing
six complex parameters $\ep_{1},\ep_{2},\ep_{3}$, $\gamma_{1},\gamma_{2},\gamma_{3}$, we take
\begin{eqnarray}
\label{eq:def_1}
E^1   & = & i\sqrt{2} \left\{ \Sigma_1\Phi^1  + \Sigma_2 (\ep_1 \Phi^1 + \ep_2 \Phi^2)\right\},\nonumber\\
E^2   & = & i\sqrt{2} \left\{ \Sigma_1\Phi^2  + \ep_3\Sigma_2  \Phi^1\right\},\nonumber\\
E^3   & = & i\sqrt{2} \left\{ \Sigma_2\Phi^3  + \Sigma_1 (\gamma_1 \Phi^3 + \gamma_2 \Phi^4)\right\},\nonumber\\
E^4   & = & i\sqrt{2} \left\{ \Sigma_2\Phi^4  + \gamma_3\Sigma_1  \Phi^3 \right\}.
\end{eqnarray}
The  $M_{(\alpha)}$ matrices take the form
\begin{equation}
M_{(1)} = \left(\begin{array}{cc} \sigma_1 + \ep_1 \sigma_2	& ~\ep_2\sigma_2	\\
				                 \ep_3 \sigma_2				& ~\sigma_1
		      \end{array}
	       \right),~~
M_{(2)} = \left(\begin{array}{cc} \gamma_1\sigma_1 + \sigma_2	&~ \gamma_2\sigma_1	\\
				                 \gamma_3 \sigma_1			&~ \sigma_2
		      \end{array}
	       \right).
\end{equation}
>From these we read off the deformed quantum cohomology relations,
\begin{eqnarray}
\sigma_1^2 + \ep_{1} \sigma_1 \sigma_2 - \ep_2 \ep_3 \sigma_2^2 & = & q_1, \nonumber\\
\sigma_2^2 + \gamma_{1} \sigma_1 \sigma_2 -\gamma_2\gamma_3 \sigma_1^2 & = & q_2,
\end{eqnarray}
in perfect agreement with~\cite{Guffin:2007mp}.

To compute correlators it is convenient to introduce the ratio $z = \sigma_2 / \sigma_1$.  The
$\sigma$ equations of motion may be re-written as
write
\begin{equation}
\sigma^2_1 = \frac{q}{s(z)},
\end{equation}
with
\begin{equation}
s(z) = 1+\ep_1 z  - \ep_2 \ep_3 z^2,
\end{equation}
and
\begin{equation}
P(z) = q_2 \det M_{(1)}(1,z) - q_1 \det M_{(2)}(1,z) =0.
\end{equation}
Let
\begin{equation}
H(\sigma_1,\sigma_2) = \det \Jt_{a,b} \det M_{(1)} \det M_{(2)}
\end{equation}
denote the measure factor.
It is easy to see that $H(\sigma_1,\sigma_2) = \sigma_1^2 H(1,z)$.
Plugging these expressions into our general formula, we find that the non-zero correlators are given by
\begin{equation}
\la \sigma_1^a\sigma_2^{2m-a} \ra =2 q_1^{m-1}\sum_{z|P(z)=0} \frac{z^{2m-a}}{s(z)^{m-1} H(1,z)}.
\end{equation}
To get explicit expressions we may use any number of simple methods---for example the couple of lines of
Maple code given in the appendix.
We find
\begin{eqnarray}
\la \sigma_1^2 \ra & = & \frac{1}{D} \left[ \ep_1 + \ep_2 \ep_3 \gamma_1\right], \nonumber\\
\la \sigma_1\sigma_2 \ra & = & \frac{1}{D} \left[ \ep_2 \ep_3 \gamma_2\gamma_3 -1 \right], \nonumber\\
\la \sigma_2^2 \ra & = & \frac{1}{D} \left[ \gamma_1 + \ep_1 \gamma_2 \gamma_3 \right],
\end{eqnarray}
with
\begin{equation}
D = (\ep_1 + \ep_2 \ep_3 \gamma_1 ) (\gamma_1 +\ep_1 \gamma_2\gamma_3) - (\ep_2\ep_3\gamma_2\gamma_3 -1)^2.
\end{equation}
These correlators agree with the results of~\cite{Guffin:2007mp}. 

There is a nice interpretation of the $q$-independent singularity $D=0$. In the Higgs phase,  $q_{1,2}\rightarrow 0$, the $\sigma$ fields are massive for  generic values of the $\ep,\gamma$  parameters, while the $\phi$ fields parametrize (up to gauge equivalence) a toric variety $V$.  The singularity at 
$D=0$ corresponds to some $\sigma$ field becoming light at some point in $V$.  This may be seen by analyzing the $|\sigma|^2 |\phi|^2$ term
in the bosonic potential.  In this example the condition for a massless $\sigma$ is the simultaneous vanishing of $\det M_{(1)}$ and
$\det M_{(2)}$ for some non-zero $\sigma$.  Solving this condition leads to $D=0$. Since the $D=0$ singularity is $q$-independent, there should be a complementary interpretation in the Coulomb phase.  Indeed, one can show that the singularity corresponds to a $\phi$ field becoming massless.  In
either case, we see that in contrast to the familiar case of the $(2,2)$ locus, the $(0,2)$ theories can exhibit mixed Higgs-Coulomb phases. 

\subsection{Resolved $\P^{4}_{1,1,2,2,2}$}
We now consider another compact toric variety with two K\"ahler parameters.  This is the resolved weighted projective
space $\P^4_{1,1,2,2,2}$.  The example is well-known from studies of Calabi-Yau hypersurfaces with $h^{1,1}=2$
\cite{Candelas:1993dm,Morrison:1994fr}. The GLSM has $n=6$ and $n-d=2$ with charge assignments
\begin{equation}
Q=\left(
\begin{array}{cccccc}
  				         0    &   0       & 1        &    1       &     1     &    1      \\
				         1    &   1       & 0        &    0        &     0     &  -2
\end{array}
\right).
\end{equation}
Although this is a different toric variety from $\P^1\times\P^1$, the massive Coulomb
analysis will not be much harder.  The counting argument given
above implies that there should be $N_E = 13$ deformations.  Although there is no obstruction to turning on all $13$ deformations, to
keep the resulting expressions simple we will only turn on three deformations that mix the $\Phi^1,\Phi^2$ fields:
\begin{eqnarray}
E^1 &=& i\sqrt{2} \left\{ \Sigma_2\Phi^1  + \Sigma_1 (\ep_1 \Phi^1 + \ep_2 \Phi^2)\right\},\cr
E^2 &=& i\sqrt{2} \left\{ \Sigma_2\Phi^2  + \Sigma_1 \ep_3 \Phi^1\right\},\cr
E^{3,4,5} &=& i\sqrt{2} \Sigma_1\Phi^{3,4,5},\cr
E^6 &=& i\sqrt{2} \left\{ \Sigma_1\Phi^6  -2 \Sigma_2 \Phi^6\right\},
\end{eqnarray}
The  $M_{(\alpha)}$ matrices take the form
\begin{eqnarray}
&&M_{(1)} = \left(\begin{array}{cc} \sigma_2 + \ep_1 \sigma_1	& ~\ep_2\sigma_1	\\
				                 \ep_3 \sigma_1				& ~\sigma_2
		      \end{array}
	       \right),~~
M_{(2)} = \diag(\sigma_1,\sigma_1,\sigma_1),\cr
&&M_{(3)} = \sigma_1 - 2\sigma_2.
\end{eqnarray}
Proceeding just as in the example of $\P^1\times\P^1$, we find
\begin{eqnarray}
&& \Jt_1 = \log\left[\frac{\det M_2 \det M_3}{q_1}\right], \qquad \Jt_2 = \log\left[\frac{\det M_1 \det M_3^{-2}}{q_2}\right],
\end{eqnarray}
which lead to the deformed quantum cohomology relations
\begin{eqnarray}
\sigma_1^3(\sigma_1 - 2\sigma_2) &=& q_1,\nonumber\\
{\sigma_2^2 + \ep_1\sigma_1\sigma_2 -\ep_2\ep_3\sigma_1^2} &=& q_2{(\sigma_1-2\sigma_2)^2}.
\end{eqnarray}
We may write these equations in terms of $\sigma_1$ and the ratio $z=\sigma_2/\sigma_1$:
\begin{eqnarray}
\sigma_1^4 &=& \frac{q_1}{(1-2z)},\cr
P(z) &=& z^2 + \ep_1 z - \ep_2\ep_3 - q_2 (1-2z)^2.
\end{eqnarray}
Plugging these expressions into the formula for the correlators, we find that the non-zero amplitudes are given by\footnote{The selection rule
$\la\sigma_1^a\sigma_2^b\ra =0$ unless $a+b=0 \mod 4$ is a simple consequence of the anomalous R-symmetry.}
\begin{eqnarray}
\langle \sigma_1^a \sigma^{4m-a}_2\rangle &=& 4q_1^{m-1}\sum_{z|P(z)=0}\frac{z^{4m-a}}{(1-2z)^{m-1} H(1,z)},
\end{eqnarray}
with the measure factor
\begin{equation}
H(1,z) = 4 (\ep_1 -4\ep_2\ep_3 + 2(1+\ep_1) z ).
\end{equation}
This form is amenable to computation.  For example, we easily compute
\begin{eqnarray}
\la \sigma_1^4 \ra & = & \frac{2}{D_1}, \cr
\la \sigma_1^3\sigma_2 \ra & = & \frac{1}{D_1},\cr
\la \sigma_1^2\sigma_2^2 \ra & = & \frac{\ep_1 -2 \ep_3\ep_3 + 2 q_2}{D_1 D_2},\cr
\la \sigma_1 \sigma_2^3 \ra & = & \frac{\ep_1^2 +\ep_2\ep_3(1-2\ep_1)+(6\ep_1-12\ep_2\ep_3+1)q_2 + 4 q_2^2}{D_1 D_2^2},
\end{eqnarray}
where
\begin{eqnarray}
D_1 & = & 1 +2 \ep_1 - 4\ep_2 \ep_3,\cr
D_2 & = & 4 q_2 -1.
\end{eqnarray}
The singularity at $D_2 = 0$ is the familiar singularity due to a quantum Coulomb branch, while the singularity at $D_1 = 0$ corresponds
to a bundle degeneration that is visible even in the large radius limit. As discussed in the previous example, this singularity has an interpretation 
as some fields becoming light. One can perform the same analysis as above, taking care to correctly account for the charges $Q_i^a$, and show 
that $D_1=0$ corresponds to a mixed Coulomb-Higgs phase. The expressions clearly show that the bundle deformations and K\"ahler parameters 
should be treated democratically.

\section{A Compact Conformal Example} \label{s:hyperexample}
There is a simple way to transform the previous example into a linear model that is expected to flow to a non-trivial SCFT in the IR.  We add
new matter multiplets $(\Phi^0,\Gamma^0)$
and take the charges to be
\begin{equation}
Q=\left(
\begin{array}{ccccccc}
  				    -4	&     0    &   0       & 1        &    1       &     1     &   ~1      \\
				    ~0	&     1    &   1       & 0        &    0        &     0     &  -2
\end{array}
\right).
\end{equation}
This GLSM is no longer compact, but we can make it compact by introducing a potential for the matter multiplets.  For example,
on the $(2,2)$ locus we may take $J_i = \p W/\p \Phi^i$, with
\begin{equation}
W  = \Phi_0 P(\Phi_1,\ldots,\Phi_5),~~~P= ( \Phi_1^8 + \Phi_2^8) \Phi_6^4 + \Phi_3^4+\Phi_4^4+\Phi_5^4.
\end{equation}
It is a simple matter to verify that when $r^{1,2} >> 0$, the low energy theory is a NLSM on the target-space described
by the smooth hypersurface $P=0$ in the resolved projective space $\P^4_{1,1,2,2,2}$.  The A-twist of this $(2,2)$ theory
was studied in detail in~\cite{Morrison:1994fr}, and we will begin by summarizing the results.

\subsection{A Review of $(2,2)$ results}
The first important observation is that this model does not possess the massive Coulomb vacua that we have been discussing
above.\footnote{On the $(2,2)$ locus this is a consequence of the non-anomalous R-symmetry---the famous
condition $\sum_i Q_i^a = 0$ for all $a$.}  In particular, the theory has no non-geometric phase, so that our techniques
for computing correlators are not immediately applicable.  However, the effective twisted superpotential is still a useful tool.
For instance, it may be used to find the singular locus of the $A$-model---i.e. the subvariety in the K\"ahler moduli space where the
SCFT is singular and correlators diverge.  In the example at hand, the twisted superpotential leads to the equations
\begin{eqnarray}
\sigma_1^3 (\sigma_1-2\sigma_2)  & = &  q_1(-4 \sigma_1)^4,\cr
\sigma_2^2 & = & q_2 (\sigma_1-2\sigma_2)^2.
\end{eqnarray}
These equations are invariant under rescaling $\sigma_1,\sigma_2$ by a constant, so that for generic
$q_1,q_2$ there is no solution, and hence the model has no non-geometric phase.  However,  at special values of $q_1,q_2$ there is
a massless $\sigma$-direction in field space, leading to a singularity in the low energy theory.  This singular locus
is determined  by computing the resultant of the $\sigma$ equations of motion.  The result is
\begin{equation}
(1-2^8 q_1)^2 -2^{18}q_1^2 q_2 = 0.
\end{equation}
As explained in~\cite{Morrison:1994fr}, this computation only gives the principal component of the singular locus.
Additional components correspond to loci where some of the gauge groups remain Higgsed, while others are in the
Coulomb phase.  In this example, there is one additional component given by $q_2 = 1/4$.

We mentioned above that since the theory lacks a non-geometric phase, we cannot directly apply our Coulomb branch
techniques to compute the correlators.  Nevertheless, there exists a way to relate the computations in this GLSM with
a non-trivial superpotential for the matter fields to computations in the toric GLSM for the ambient variety.  This is given
by the quantum restriction formula derived in~\cite{Morrison:1994fr}:
\begin{equation}
\lad \sigma_1^a \sigma_2^b \rad = \la  \sigma_1^a \sigma_2^b\frac{-K}{1-K} \ra,
\end{equation}
where $\lad\cdots\rad$ denotes correlators on the hypersurface, and
\begin{equation}
-K = \sum_{i>0} Q_i^a\sigma_a=4\sigma_1
\end{equation}
is the operator corresponding to the anti-canonical class of the hypersurface.  In a sense, it should not be a surprise that
such an expression should exist:  the $A$-model is invariant under small changes of the coefficients in the chiral
superpotential, so that it is reasonable that the correlators would only depend on coarse data like the anti-canonical
class of the surface.  What is perhaps surprising is the elegant form that the relation takes. Given this relation, we may
use the Coulomb branch techniques in the GLSM for the ambient variety to compute correlators in the GLSM corresponding
to the Calabi-Yau three-fold~\cite{Melnikov:2006kb}.   This is a significant simplification, as the quantum restriction formula
typically requires the evaluation of an infinite number of correlators in the ambient GLSM.

\subsection{$(0,2)$ Deformations}
How do we expect these results to change off the $(2,2)$ locus?  First, since the $J_i$ are non-zero,
there is a non-trivial requirement for $(0,2)$ SUSY:  $\sum_i E^i J_i = 0$.  Deformations of the $E^i$ will, in general,
need to be accompanied by deformations of the $J_i$.  In the case of the $\ep_{1,2,3}$ deformations considered above,
it is sufficient to deform $J_0$:
\begin{equation}
\Delta J_0 = 2 (\ep_1 \Phi_1^8 + \ep_2 \Phi_1^7 \Phi_2+\ep_3 \Phi_1 \Phi_2^7)\Phi_6^4.
\end{equation}
The general arguments of~\cite{Silverstein:1995re,Basu:2003bq,Beasley:2003fx} suggest that these deformations should
correspond to marginal deformations of the $(0,2)$ SCFT.

We still expect the analysis of the effective potential for the $\Sigma_a,\Upsilon_a$ to hold.  After all, when the $\sigma_a$
have large VeVs, we expect the $J_i$ couplings to be unimportant.  Using our formula for the potential, we arrive at
the $\sigma$ equations of motion:
\begin{eqnarray}
\sigma_1^3 (\sigma_1 - 2\sigma_2) & = & q_1  (-4\sigma_1)^4, \cr
\det M_{(1)} & = & q_2(\sigma_1-2\sigma_2)^2.
\end{eqnarray}
These are again homogeneous in the $\sigma_a$, so that the common solutions exist only when
\begin{equation}
D_\ep = (1-2^8 q_1)^2 - 2^{18} q_1^2 q_2 + 2 \ep_1(1-2^8 q_1) - 4 \ep_2 \ep_3 = 0.
\end{equation}
This is the $(0,2)$ deformation of the principal component of the singular locus described above.  In particular, we see
that even in the $q_1, q_2 \to 0$ limit it is possible to get singularities by degenerating the bundle structure.  While not surprising
in general, it is gratifying to have such a concrete realization of this phenomenon.  There should be no difficulty in generalizing the
analysis to find other components of the singular locus.  In the example at hand it is easy to see that the
$q_2 = 1/4$ component remains undeformed.

The generalization of the quantum restriction formula is more subtle.  In the $A$-model there was a simple
reason for the decoupling of the matter superpotential:  it corresponded to $Q_T$-exact deformations of the topological theory.  In
the $(0,2)$ half-twisted theory this is not so clear.  After all, the F-I terms as well as the $J_i$ are just terms in a $(0,2)$ superpotential, and
there does not seem to be an obvious reason that the half-twisted correlators of the $\sigma_a$ should not depend on the coefficients
in the $J_i$.  However, there may well exist a modification of the Morrison-Plesser quantum restriction formula that will compute
the dependence on all the moduli.  We plan to return to finding a suitable modification in future work.  In what follows, we will content
ourselves with presenting some evidence for such a formula.

To test our expectations of quantum restirction, let us naively apply the usual Morrison-Plesser formula to the $(0,2)$ correlators.
The change in the $J_0$ simply made a small deformation in the hypersurface, so the anti-canonical class certainly remains unchanged.
Inserting $-K = 4\sigma_1$, we find
\begin{equation}
\lad \sigma_1^{3-a} \sigma_2^a \rad = \la \frac{4 \sigma_1^{4-a} \sigma_2^a}{1-4^4 \sigma_1^4} \ra
=  \sum_{z|P(z)} G(z),
\end{equation}
where
\begin{equation}
G(z) =  \frac{16 z^a (1-2z)}{H(1,z) (1-2 z -4^4 q_1)}.
\end{equation}
The sum is easy to evaluate by writing it as a contour integral around the zeroes of $P$, and then pulling the contour off onto the other
poles.  We find
\begin{equation}
\lad \sigma_1^{3-a} \sigma_2^a \rad = -
\left\{ \Res_{z=\frac{1-4^4 q_1}{2}}
+ \Res_{z= \frac{4\ep_2\ep_3 -\ep_1}{2(1+\ep_1)}} + \Res_{z=\infty} \right\} \frac{G(z)P'(z)}{P(z)}.
\end{equation}
Using this expression, the three-point functions are given by
\begin{eqnarray}
\lad \sigma_1^3 \rad 			& = & \frac{8}{D_\ep}, \cr
\lad \sigma_1^2 \sigma_2 \rad	& = & \frac{4(1-2^8 q_1)}{D_\ep}, \cr
\lad \sigma_1 \sigma_2^2 \rad	& = & \frac{4(2^{10} q_1 q_2 -2 q_2 +2^8 \ep_1q_1+2\ep_2\ep_3 -\ep_1)}{(1-4q_2)D_\ep}, \cr
\lad  \sigma_2^3 \rad 		& = & 4\left[q_2(1+4q_2 - 2^8 q_1 - 3072 q_1 q_2) + \ep_1^2(1-2^8 q_1)  \right.\cr
&~&~~
+2\ep_1 (-2^{10} q_1 q_2 + 3 q_2 -\ep_2\ep_3) \cr
&~&~~ \left.
 +\ep_2\ep_3(- 2^8 q_1  + 2^{10} q_2 q_1 + 1 - 12 q_2)\right]/(1-4q_2)^2 D_\ep.
\end{eqnarray}
Remarkably, the correlators in the ambient GLSM sum up to produce the singularities expected from the Coulomb branch
analysis of the model with hypersurface!  We take this to be a good indication that the dependence on the $J_i$ parameters
may well be simple enough to be captured by a suitable modification of the quantum restriction formula.

\section{Conclusions} \label{s:conclusions}
We have developed a simple method to compute an interesting set of correlators in half-twisted compact, toric sigma models.
These correlators should be interpretable as the $(0,2)$ generalization of genus zero Gromov-Witten invariants described in
\cite{Katz:2004nn}. A by-product of our analysis was a simple derivation of the deformed quantum cohomology relations in these
theories.  We have applied our results to several examples, and it is fairly clear that they should apply to more intricate models
without excessive computational burden.

Clearly, the most interesting extensions of this work lie in applications to theories of the sort considered in our last example:
$(0,2)$ deformations of compact linear models that flow to non-trivial SCFTs in the IR.  Heartened by our results for the
compact toric theories, we would like to study the modification to the quantum restriction formula in more detail.
Once this aspects of the problem is well understood, we will be in a much better position to develop
$(0,2)$ generalizations of special geometry and mirror symmetry in these deformed theories.  Another exciting direction would
be to attempt to apply these methods to $(0,2)$ theories without a $(2,2)$ locus.  We hope to report on these matters in the
near future.

%%%%%%%%%%%%%%%%%%%%%%%%%%%%%%%%%%%%%%%
\acknowledgments
We would like to thank J.~Guffin for useful correspondence and S.~Sethi for comments on the manuscript.  I.M. would like to
thank SLAC and the Stanford Institute for Theoretical Physics for hospitality while some of this work was completed.  J.M. is
supported by the Ledley Fellowship.  This article is based upon work supported in part by the National Science Foundation
under Grants PHY-0094328 and PHY-0401814 .

\appendix
\section{A Maple Routine to Compute Correlators} \label{a:code}
Below we present a simple Maple code that computes correlators in the $\P^1\times \P^1$ model.
We hope the simplicity of this routine makes it clear that our results lead to correlators with a minimum
of computational effort.  Lines beginning with \# are comments.
\begin{verbatim}
with(LinearAlgebra):

#The deformations.  s[1], s[2] are the sigma fields.
#To compare to Guffin and Katz, set a[4]=b[4]=0
#and remaining a[i] = epsilon[i], b[i] = gamma[i].

M[1] := Matrix([ [s[1]+a[1]*s[2], a[2]*s[2]        ],
               [   a[3]*s[2]    , s[1]+a[4]*s[2]   ]
             ]);
M[2] := Matrix([ [s[2]+b[1]*s[1], b[2]*s[1]        ],
               [   b[3]*s[1]    , s[2]+b[4]*s[1]   ]
             ]);

DM[1]:= Determinant(M[1]);
DM[2]:= Determinant(M[2]);

#The J's
J[1] := log (DM[1]/q[1]);
J[2] := log (DM[2]/q[2]);

#The second derivatives of J
JJ := Matrix(2,2):
JJ[1,1] := diff( J[1], s[1]);
JJ[1,2] := diff( J[1], s[2]);
JJ[2,1] := diff( J[2], s[1]);
JJ[2,2] := diff( J[2], s[2]);

DJJ := Determinant(JJ);

#the combined measure.
H := simplify(DJJ*DM[1]*DM[2]);

#substitution of s[2] = z s[1].
HS := collect(subs(s[1]=1,s[2]=z, H),z);
S := collect(subs(s[1]=1,s[2]=z,DM[1]),z);

#equation for z
P := simplify(subs(s[1]=1,s[2]=Z, DM[1]*q[2]-DM[2]*q[1]));

#Non-Zero Correlators have a+b = 2m.
#the factor of 2 follows from summing over
#solutions for s[1]:
C := (a, m) ->
    simplify( 2*q[1]^(m-1)
             *sum(z^(2*m-a)*S^(1-m)*HS^(-1),
                  z=RootOf(P,Z)
                 )
            );
\end{verbatim}

%\bibliographystyle{utphys}
%\bibliography{refs}

\begin{thebibliography}{10}

\bibitem{Morrison:1994fr}
D.~R. Morrison and M.~Ronen~Plesser, ``Summing the instantons: Quantum
  cohomology and mirror symmetry in toric varieties,'' {\em Nucl. Phys.} {\bf
  B440} (1995) 279--354,
\href{http://arXiv.org/abs/hep-th/9412236}{{\tt hep-th/9412236}}.
%%CITATION = HEP-TH/9412236;%%.

\bibitem{Witten:1993yc}
E.~Witten, ``Phases of {N} = 2 theories in two dimensions,'' {\em Nucl. Phys.}
  {\bf B403} (1993) 159--222,
\href{http://arXiv.org/abs/hep-th/9301042}{{\tt hep-th/9301042}}.
%%CITATION = HEP-TH/9301042;%%.

\bibitem{Hori:2000kt}
K.~Hori and C.~Vafa, ``Mirror symmetry,''
\href{http://arXiv.org/abs/hep-th/0002222}{{\tt hep-th/0002222}}.
%%CITATION = HEP-TH/0002222;%%.

\bibitem{Melnikov:2006kb}
I.~V. Melnikov and M.~R. Plesser, ``A-model correlators from the {C}oulomb
  branch,'' {\em JHEP} {\bf 02} (2006) 044,
\href{http://arXiv.org/abs/hep-th/0507187}{{\tt hep-th/0507187}}.
%%CITATION = JHEPA,0602,044;%%.

\bibitem{Distler:1986wm}
J.~Distler, ``Resurrecting (2,0) compactifications,'' {\em Phys. Lett.} {\bf
  B188} (1987)
431--436.
%%CITATION = PHLTA,B188,431;%%.

\bibitem{Silverstein:1995re}
E.~Silverstein and E.~Witten, ``Criteria for conformal invariance of (0,2)
  models,'' {\em Nucl. Phys.} {\bf B444} (1995) 161--190,
\href{http://arXiv.org/abs/hep-th/9503212}{{\tt hep-th/9503212}}.
%%CITATION = HEP-TH/9503212;%%.

\bibitem{Basu:2003bq}
A.~Basu and S.~Sethi, ``World-sheet stability of (0,2) linear sigma models,''
  {\em Phys. Rev.} {\bf D68} (2003) 025003,
\href{http://arXiv.org/abs/hep-th/0303066}{{\tt hep-th/0303066}}.
%%CITATION = HEP-TH/0303066;%%.

\bibitem{Beasley:2003fx}
C.~Beasley and E.~Witten, ``Residues and world-sheet instantons,'' {\em JHEP}
  {\bf 10} (2003) 065,
\href{http://arXiv.org/abs/hep-th/0304115}{{\tt hep-th/0304115}}.
%%CITATION = HEP-TH/0304115;%%.

\bibitem{Adams:2003zy}
A.~Adams, A.~Basu, and S.~Sethi, ``(0,2) duality,'' {\em Adv. Theor. Math.
  Phys.} {\bf 7} (2004) 865--950,
\href{http://arXiv.org/abs/hep-th/0309226}{{\tt hep-th/0309226}}.
%%CITATION = HEP-TH/0309226;%%.

\bibitem{Sharpe:2005fd}
E.~Sharpe, ``Notes on correlation functions in (0,2) theories,''
\href{http://arXiv.org/abs/hep-th/0502064}{{\tt hep-th/0502064}}.
%%CITATION = HEP-TH/0502064;%%.

\bibitem{Adams:2005tc}
A.~Adams, J.~Distler, and M.~Ernebjerg, ``Topological heterotic rings,'' {\em
  Adv. Theor. Math. Phys.} {\bf 10} (2006) 657--682,
\href{http://arXiv.org/abs/hep-th/0506263}{{\tt hep-th/0506263}}.
%%CITATION = HEP-TH/0506263;%%.

\bibitem{Witten:1991zz}
E.~Witten, ``Mirror manifolds and topological field theory,''
\href{http://arXiv.org/abs/hep-th/9112056}{{\tt hep-th/9112056}}.
%%CITATION = HEP-TH/9112056;%%.

\bibitem{Melnikov:2007xi}
I.~V. Melnikov and S.~Sethi, ``Half-twisted (0,2) {L}andau-{G}inzburg models,''
\href{http://arXiv.org/abs/arXiv:0712.1058 [hep-th]}{{\tt arXiv:0712.1058
  [hep-th]}}.
%%CITATION = ARXIV:0712.1058;%%.

\bibitem{Katz:2004nn}
S.~H. Katz and E.~Sharpe, ``Notes on certain (0,2) correlation functions,''
  {\em Commun. Math. Phys.} {\bf 262} (2006) 611--644,
\href{http://arXiv.org/abs/hep-th/0406226}{{\tt hep-th/0406226}}.
%%CITATION = HEP-TH/0406226;%%.

\bibitem{Guffin:2007mp}
J.~Guffin and S.~Katz, ``Deformed quantum cohomology and (0,2) mirror
  symmetry,''
\href{http://arXiv.org/abs/arXiv:0710.2354 [hep-th]}{{\tt arXiv:0710.2354
  [hep-th]}}.
%%CITATION = ARXIV:0710.2354;%%.

\bibitem{Cox:2000vi}
D.~Cox and S.~Katz, {\em Mirror symmetry and Algebraic Geometry}.
\newblock AMS, 2000.

\bibitem{Hori:2003ic}
K.~Hori {\em et al.}, {\em Mirror Symmetry}.
\newblock AMS, 2003.

\bibitem{Vafa:1990mu}
C.~Vafa, ``Topological {L}andau-{G}inzburg models,'' {\em Mod. Phys. Lett.}
  {\bf A6} (1991)
337--346.
%%CITATION = MPLAE,A6,337;%%.

\bibitem{Candelas:1993dm}
P.~Candelas, X.~De~La~Ossa, A.~Font, S.~H. Katz, and D.~R. Morrison, ``Mirror
  symmetry for two parameter models. {I},'' {\em Nucl. Phys.} {\bf B416} (1994)
  481--538,
\href{http://arXiv.org/abs/hep-th/9308083}{{\tt hep-th/9308083}}.
%%CITATION = HEP-TH/9308083;%%.

\end{thebibliography}

\providecommand{\href}[2]{#2}\begingroup\raggedright\endgroup

\end{document}